
\magnification=\magstep1
\font\huge=cmr10 scaled \magstep2
\def\*{^}
\overfullrule=0pt
\def\h{h\*\vee} \def\c{\chi} \def\C{\chi\* *}

\def\L{\Lambda}   \def\la{\lambda}
 \def\Z{{\bf Z}}
       \def\g{{\hat g}}
\def\equi{\,{\buildrel \rm def \over =}\,}
\def\sp{\;}    \def\eg{{\it e.g.}$\sp$} \def\ie{{\it i.e.}$\sp$}
 \def\T{{\cal T}}
\def\i{\item} \def\eps{\epsilon}
\def\sp{\,\,\,}
\def\d{\delta}
\def\CIZ{[1]} \def\MS{[2]} \def\SY{[3]}  \def\CMS{[1,2,3]} \def\ITZ{[4]}
\def\COMM{[5]}  \def\ITCO{[4,5]}
\def\KA{[6]} \def\CIK{[1,6]} \def\Kas{[7]} \def\GA{[8]}
\def\CL{[10]} \def\HET{[11]}
\def\WA{[12]} \def\WAH{[12,11]}  \def\PAR{[5,16,8]}
\def\HK{[13]}
\def\CS{[14]}\def\Gal{[15]}\def\GHF{[13,19]}
\def\RTW{[16]}  \def\KAC{[17]}
\def\CG{[18]} \def\SUSU{[9]}
\def\FKSV{[19]}  \def\CE{[20]} \def\BOW{[21]} \def\SCH{[22]}

{ \nopagenumbers
\rightline{January, 1994}\bigskip\bigskip
\centerline{{\bf \huge The Rank Four Heterotic Modular Invariant}}
\bigskip\centerline{{\bf \huge Partition Functions}}
\bigskip \bigskip
\centerline{Terry Gannon}
\centerline{{\it Institut des Hautes Etudes Scientifiques}}
\centerline{{\it 91440 Bures-sur-Yvette, France}}\bigskip
\centerline{Q. Ho-Kim}
\centerline{{\it Physics Department, Laval University}}
\centerline{{\it Sainte-Foy, Que., Canada G1K 7P4}}\bigskip
\bigskip \bigskip \centerline{{\bf Abstract}}\bigskip

In this paper, we develop several general techniques to investigate modular
invariants of conformal field theories whose algebras of the holomorphic
and anti-holomorphic sectors are different.  As an application, we find all
such ``heterotic'' WZNW physical invariants of (horizontal) rank four: there
are exactly seven of these, two of which seem to be new.  Previously, only
those of rank $\le 3$ have been completely classified.  We also find all
physical modular invariants for $su(2)_{k_1}\times su(2)_{k_2}$, for
$22>k_1>k_2$, and $k_1=28$, $k_2<22$, completing the classification of ref.{}
\SUSU.
\vfill \eject }


\pageno=1
\bigskip \centerline{{\bf 1. Introduction}} \bigskip

In the study of two-dimensional conformal field theories, the partition
functions play an important role. In particular, they could give us some
insight into the general solution of the broader and as yet largely unsolved
problem of classifying all such theories. Unfortunately, only in a few special
cases are these functions completely known; much work remains to be done.

For a two-dimensional conformal field theory that has an operator algebra
decomposable into a pair of commuting holomorphic and anti-holomorphic chiral
algebras, ${\g_L}$ and ${\g_R}$, and a space of states which can be written
as a finite sum of irreducible representations $(\la_L,\la_R)$ of
${\g_L\times\g_R}$ with multiplicity $ N_{\la_L \la_R}$, the partition
functions are combinations of bi-products of characters $\c_{\la _L}$
and $\c_{\la_R}$ of the form

   $$Z=\sum N_{\la_L \la_R} \c_{ \la_L}\,\C_{\la_R}. \eqno(1.1)$$

To classify all partition functions of a given algebra $(\g_L, \g_R)$
is to find all combinations (1.1) such that: (P1) $Z$ is invariant under
transformations of the modular group; (P2) all the coefficients $N_{ \la_L
\la_R}$ are non-negative integers; and (P3) the vacuum exists and is
non-degenerate, that is, $N_{11}=1$, with $\la=1$ denoting the vacuum.

Any function $Z$ satisfying the modular invariance condition (P1)  will be
called an {\it invariant\/}; if in addition it satisfies the condition that
every coefficient $N_{\la_L\la_R}\geq 0$, then it is said to be a {\it positive
invariant\/}; and finally, if the conditions (P1), (P2) and (P3) are all
met, then it is considered to be a {\it physical invariant\/}.  Clearly
any conformal theory must have {\it at least\/} these three properties to be
physically meaningful, but often additional conditions are imposed.

In the past, most authors have assumed further that the algebras and levels of
the holomorphic and anti-holomorphic sectors of the theory are identical
($\g_L=\g_R$ and $k_L=k_R$); when it is necessary to emphasize this restriction
we will use the qualitative {\it `non-heterotic'\/} or {\it `symmetric'}.  In
the past decade, extensive work has been done in this direction, and has been
reported in refs.{} \CMS, among others.  But classification proofs, which
determine all the physical invariants that belong to a certain class, exist
only for a few cases.  Namely: at level one, the simple Lie algebras
$A_n\*{(1)}$, $B_n\*{(1)}$, $C_n\*{(1)}$, $D_n\*{(1)}$, and the five
exceptional algebras \ITCO; and, at an arbitrary level, the untwisted Kac-Moody
algebra $A_1\*{(1)}$ (see \CIK) and the coset models based on it, such as the
minimal unitary Virasoro models \CIZ{} and the  $N=1$ minimal superconformal
models \Kas, and, also, the algebra $A_2\*{(1)}$ (ref.{} \GA). Finally, thanks
in part to work done in this paper, the $A_1+A_1$ classification has now been
completed for all levels \SUSU.

In the present paper we will address rather the {\it heterotic\/} case, where
the (affine) algebras of the holomorphic and anti-holomorphic sectors of the
theory are different, $(\g_L,k_L)\neq (\g_R,k_R)$. We will focus on the rank
four case, by which we mean rank($\bar{g_L})+$rank($\bar{g_R}$)=4.  However our
methods will be of value for higher ranks as well.

Relatively little work has been done in this direction (one of the few recent
exceptions is ref.{} \CL) although, if the heterotic string model is any
indication, heterotic invariants could lead to pheno\-meno\-logically
interesting models. The heterotic invariants of rank$\le 3$ were found in
ref.{} \HET:
$$\eqalignno{{\rm for}\sp(A_{2,1};A_{1,4}):\quad
Z=&\chi_{11}\{\chi_1+\chi_5\}\**+
\{\chi_{21}+\chi_{12}\}\chi_3\**;&(1.2a)\cr
{\rm for}\sp(C_{2,1};A_{1,10}):\quad Z=&\chi_{11}\{\chi_1+\chi_7\}\**+
\chi_{21}\{\chi_4+\chi_8\}\**&\cr
&+\chi_{12}\{\chi_5+\chi_{11}\}\**;&(1.2b)\cr
{\rm for}\sp(G_{2,1};A_{1,28}):\quad Z=&\chi_{11}\{\chi_1+\chi_{11}+
\chi_{19}+\chi_{29}\}\**&\cr
&+\chi_{12}\{\chi_7+\chi_{13}+\chi_{17}+\chi_{23}\}\**;&(1.2c)\cr
{\rm for}\sp(A_{1,3}A_{1,1};A_{1,28}):\quad Z=&\{\chi_{1}\chi_1+\chi_4\chi_2\}
\cdot\{\chi_1+\chi_{11}+\chi_{19}+\chi_{29}\}\**&\cr
&+\{\chi_{2}\chi_2+\chi_{3}\chi_1\}\cdot\{\chi_7+\chi_{13}+\chi_{17}
+\chi_{23}\}\**;&(1.2d)\cr}$$
where the subscripts of the characters are the Dynkin labels of
the highest weight$+\rho$ ($\rho=$sum of the fundamental weights).

The heterotic cases differ from the more familiar symmetric cases in that they
are considerably rarer: for any algebra $\g_L=\g_R$ and level $k_L=k_R$, there
is at least one physical invariant (namely the diagonal one); however, due to
severe constraints, the existence of a physical invariant when $(\g_L,k_L)\neq
(\g_R,k_R)$ is quite exceptional.

Heterotic invariants can be found using conformal embeddings, though not all
heterotic invariants can be obtained in this manner. Given any heterotic
invariant, others can be generated using simple currents.  Simple currents can
also take a symmetric invariant to an invariant whose maximally extended left
and right chiral algebras are no longer isomorphic; however, since simple
currents do not affect the underlying affine algebra, these invariants are not
``heterotic'' in our definition of the word.

In sect.~2 we begin by finding all solutions to the constraint that the
conformal charges $c_L$ and $c_R$ be equal (mod 24). There are infinitely many
solutions corresponding to $g_L=g_R=A_1+A_1$ (we will call these the
``AA-types''), but only 61 corresponding to all other total rank 4 algebras
(which we will call the ``non-AA-types''). They are listed in Table 1.  We then
introduce the concept of ``null augments'' and use it to describe a parity
test. This test will reduce the number of non-AA-types we have to consider from
61 to 14. These 14 are listed in Table 2.  In sect.~3 we discuss the lattice
method \WAH, which we use to find all physical heterotic invariants for some of
the 14 non-AA-types. In sect.~4 another test, based on \MS, is considered; it
is used to eliminate all but one of the AA-types.  In sect.~5 we handle all
remaining types. The complete list of all heterotic rank four partition
functions is given in sect.~6; there are precisely seven of them.  The Appendix
presents the results of a computer search (using methods developed in \HK) for
the symmetric $A_1+A_1$ physical invariants at levels $k_1\neq k_2$. These
results are new (formerly only those for $k_1=k_2$ were known), and are
exploited in sect.~4. They also complete the $A_1+A_1$ classification.

\bigskip \bigskip
		 \centerline{{\bf 2. The candidate types }}
\bigskip\nobreak

In this section we give two strong tests [see eqs.~(2.2) and (2.10$b$)] which
heterotic types must pass in order for heterotic invariants to exist. Of
course, both conditions are automatically satisfied by symmetric types.

The results of this section are summarized in Table 1. First a few remarks
about our notation.

By the {\it type} ${\cal T}$ of a physical invariant we mean a list of
(left and right) algebras and levels with the left-moving (\ie holomorphic)
and the right-moving (\ie anti-holomorphic) sectors separated by a semi-colon:
$$\eqalignno{
{\cal T}=&({\cal T}_L;{\cal T}_R)=(g_{L1,k_{L1}}\cdots
g_{Ll,k_{Ll}};g_{R1,k_{R1}}\cdots g_{Rr,k_{Rr}})&\cr
=&(\{g_{L1},k_{L1}\},\ldots;\ldots,\{g_{Rr},k_{Rr}\}).&(2.1)\cr}$$
By a {\it positive type} we mean a type with all levels $k_{Li},k_{Rj}>0$.
By a {\it null type} we mean a type whose levels satisfy
$k_{L1}=\cdots=k_{Rr}=0$.
Because the character of a level zero affine algebra is identically equal to 1,
null types do not contribute to the partition function, so for most purposes
can be avoided. They are, however, necessary for parity calculations (see
subsect.~2.3 below) and the lattice method (see sect.~3 below). Positive types
will be generally denoted by ${\cal T}\*+$, and null ones by ${\cal T}\*0$.

\bigskip
	 \noindent{{\it 2.1 The central charge condition}}
\medskip\nobreak
Most types ${\cal T}\*+$ cannot be realized by a heterotic physical invariant
because of a number of stringent conditions. The most obvious condition
constraining ${\cal T}\*+$ is that the central charges of left-(right-)moving
sectors satisfy $c_L\equiv c_R$ (mod 24), or more explicitly $$
\sum_{i=1}\*l {\rho_{Li}\*2\over k_{Li}'}-\sum_{j=1}\*r{\rho_{Rj}\*2
\over k'_{Rj}}\equiv \sum_{i=1}\*l {\rho_{Li}\*2\over \h_{Li}}-
\sum_{j=1}\*r{\rho_{Rj}\*2\over \h_{Rj}}\sp ({\rm mod} \sp 2),\eqno(2.2)
$$
where $k'_{Li}=k_{Li}+\h_{Li}$ and $k'_{Rj}=k_{Rj}+\h_{Rj}$ are the heights.
The notation used here is quite standard and is explicitly described in \eg
\HET; in particular, $\rho$ is the sum of the fundamental weights and
$\h$ the dual Coxeter number.

Eq.~(2.2) must be satisfied by any type that has a modular invariant $N$
satisfying $N_{\rho_L,\rho_R}\neq 0$. In particular, it must be satisfied by a
physical invariant of any type.
In this subsection we find all solutions $\T\*+$ to eq.~(2.2) that are of rank
4 (\ie 4=$\sum n_{Li}+\sum n_{Rj}$, where $n_{Li}=rank(g_{Li}), n_{Rj}=
rank(g_{Rj})$). Any such type will be called a {\it candidate}.

A useful fact is the following:
\item{$(*)$} Any solution $x,y\in \Z$ to
$$y={ax\over a+bx},\eqno(2.3a)$$
where $ a=\prod_{i=1}\*mp_i\*{a_i}$ is the prime decomposition of $a$, and
where $gcd(a,b)=1$, also satisfies
$$x={\pm\prod_{i=1}\*m p_i\*{b_i}-a\over b},\sp{\rm where}\sp 0\le b_i\le 2a_i.
\eqno(2.3b)$$

This statement is proved simply by finding bounds for the powers of the prime
divisors of the numerator and denominator of eq.~(2.3$a$).

The above rule permits (2.2) to be solved for the heights, given the algebras
$g_{Li}, g_{Rj}$, whenever only two heights are unknown. When there are more
than two independent heights, the strategy is to use (2.2) to bound from above
all but two of the heights, and then to apply ($*$). When this strategy is
successful, only finitely many solutions to (2.2) will exist (for the given
choice of $g_{Li},g_{Rj}$), and $(*$) will permit these to be enumerated. When
this strategy is not successful (\ie there remain at least three unbounded
heights), there will be an infinite family of solutions to (2.2). For the rank
4 case we are considering here, this strategy is successful for all possible
choices of $g_{Li},g_{Rj}$, except for one: $g_L=g_R=A_1+A_1$. In that case
(2.2) becomes $${1\over k_1'}+{1\over k_2'}={1\over k_3'}+{1\over k_4'},
\eqno(2.4)$$ making a slight change of notation (the `$\equiv$' in (2.2)
becomes an `=' in (2.4) because both sides lie between $\pm 1/3$). We will
defer the analysis of this infinite class of candidates, which we call the
AA-types, until sect.~4.

We now explicitly give an example of how to solve (2.2) for a non-AA-type.
Consider $g_L=C_2$ and $g_R=A_1+A_1$, then eq.~(2.2) becomes $${15-2k_L'\over
3k_L'}={1\over k_{R1}'}+{1\over k_{R2}'},\eqno(2.5a)$$ from which we derive
$3<k_L'<7.5$. Choosing now $k_L'=7$, say, gives $$k_{R2}'={21k_{R1}'\over
k_{R1}'-21},\eqno(2.5b)$$ and now eq.(2.3$b$) tells us $k_{R1}'=21+3\*i\cdot
7\*j$ for $0\leq i,j\leq 2$, leading to 5 different candidates, namely the
types 40-44 in Table 1.

The list of all non-AA-types is given in the second column of Table 1.
The table also includes all AA-types (there is only one) which survive the
cardinality test described below in sect.~4.

\bigskip \noindent{{\it 2.2 Null augments}}\medskip

As mentioned earlier in this section, null types do not contribute to the
partition function (1.1) and so for most purposes can be ignored. However they
do have two related applications to the classification of heterotic invariants.
One is the lattice method for constructing invariants, discussed in the
following section. The other is an extremely useful relation between different
coefficients of any modular invariant.  We shall call it the {\it parity rule},
and discuss it in the following subsection. For both these reasons, it will be
convenient to define null augments.

First, let $\T=(\T_L;\T_R)$ be any type. Let $M_{Li}$ be the coroot lattice
of $g_{Li}$. By $M(\T_L)$ we mean the scaled coroot lattice
$$M(\T_L)=\bigl( \sqrt{k_{L1}'}M_{L1}\bigr)\oplus\cdots\oplus\bigl( \sqrt{
k_{Ll}'}M_{Ll}\bigr).$$
So $M(\T_L)$ will be a lattice of total dimension $n_L=\sum n_{Li}$.
Define $M(\T_R)$ by a similar expression.

Now, let $\T'$ and $\T''$ be any two types. By their {\it augment}
$\T=\T'+\T''$ we simply mean the concatenation of the two types (so \eg
$M(\T_L)=M(\T'_L)\oplus M(\T''_L)$).

Suppose we are given a positive type $\T\*+=(\T\*+_L;\T\*+_R)$ satisfying
(2.2). By its {\it null augment} \HET{} we mean any null type $\T\*0
=(\T\*0_L;\T\*0_R)$ for which the augment $\T=\T\*++\T\*0$ satisfies the
following two equations:
$$\eqalignno{M(\T_L)&\sim M(\T_R);&(2.6a)\cr
\sum_{i=1}\*l {\rho_{Li}\*2\over \h_{Li}}-\sum_{j=1}\*r{\rho_{Rj}\*2
\over \h_{Rj}}&\equiv \sum_{i=1}\*l {n_{Li}\over 4}-
\sum_{j=1}\*r{n_{Rj}\over 4}\sp ({\rm mod} \sp 2).&(2.6b)}$$
`$\sim$' in (2.6$a$) denotes the lattice
{\it similarity} relation \Gal, which
is closely related to rational equivalence. Its geometrical significance
here will be briefly discussed in sect.~3. Among other things it requires
that the product $|M(\T_L)|\,|M(\T_R)|$ of determinants be a perfect square.
The sums in (2.6$b$) are defined over the whole augment $\T$.

There will always be infinitely many different null augments for any candidate
\HET, but which one is chosen will not affect any subsequent calculations. In
Table 1 we list one choice for each candidate.

The task of finding a null augment for a given candidate is not difficult.  In
\Gal{} a calculus is developed for determining whether two lattices are
similar. We will do an example here, verifying (2.6$a$) is satisfied for
candidate 11 of Table 1.

Here, $\T_L=(A_{2,3}C_{2,0}A_{1,0})$ and $\T_R=(A_{1,2}A_{1,10}A_{3,0})$,
so $M(\T_L)=\bigl(\sqrt{6}A_2\bigr)\oplus\bigl(\sqrt{6}\,\Z\*2\bigr)\oplus
\bigl( \sqrt{4}\,\Z\bigr)$ and $M(\T_R)=\bigl(\sqrt{8}\,\Z\bigr)\oplus
\bigl(\sqrt{24}\,\Z \bigr)\oplus\bigl(\sqrt{4}A_3\bigr)$. We will use the
convenient abbreviation
$\{m_1,\ldots,m_\ell\}$ for the orthogonal lattice $(\sqrt{m_1}\,\Z)\oplus
\cdots\oplus (\sqrt{m_\ell}\,\Z)$. Since $A_2\sim\{3,3,3,1,1,1\}$ and
$A_3\sim\{1,1,1\}$, we get $\sqrt{6}A_2\sim\{18,18,18,6,6,6\}\sim\{3\}$
and $\sqrt{4}A_3\sim\{4,4,4\}\sim\{1\}$, using the $\sim$-calculus. Also,
$\{6,6,4\}\sim\{3,3\}$ and  $\{8,24\}\sim\{3,3,3\}$, so we get $M(\T_L)\sim
\{3,3,3\}\sim M(\T_R)$, and thus (2.6$a$) holds.

One thing should be mentioned. Any number of copies of the null type
$(A_{1,0})$  can be added to either side of a null augment, producing another
null augment. The only relevant change is to the dimensions $n_L,n_R$.  In
particular, for some purposes it will be most convenient (see sect.~3) to
choose the null augment so that $\T$ satisfies $$n_L\equiv n_R\sp ({\rm mod}\sp
8),\eqno(2.7a) $$ while for other purposes it will be most convenient (see
subsect.~2.3) to choose the null augment so that $\T$ satisfies $$n_L\equiv
n_R\sp ({\rm mod}\sp 2). \eqno(2.7b)$$ The choices in Table 1 all satisfy
$(2.7b)$. In some cases this was accomplished by including a copy of $A_{1,0}$.

\bigskip \noindent{{\it 2.3 The parity rule}}\medskip

In the classification of non-heterotic physical invariants, one of the most
powerful tools is the {\it parity rule} \PAR. In this subsection
we will obtain its heterotic counterpart, and use it to formulate a strong
condition a type must satisfy if it is to be realized by a heterotic physical
invariant.

First we need a few remarks about the weights and Weyl group of affine
algebras. Let $g$ be any finite dimensional Lie algebra, let $M$ denote its
coroot lattice, and choose any non-negative level $k$. The set of all weights
of $g$ is just the dual lattice $M\**$. Define the sets $P_+(g,k+\h)$ and
$P_{++}(g,k+\h)$ by
$$\eqalignno{
P_+(g,k+\h)=&\{\la\in M\**\,|\,\la_i\geq 0,\,\,\sum_{i=1}\*n
a_i\*\vee\la_i\leq k+\h\},&(2.8a)\cr
P_{++}(g,k+\h)=&\{\la\in M\**\,|\,\la_i> 0,\,\,\sum_{i=1}\*n
a_i\*\vee\la_i< k+\h\}.&(2.8b)\cr}$$
where $n$ is the rank of $g$ and the $a_i\*\vee$ are positive integers called
the {\it colabels} \KAC{} of the affinization $g\*{(1)}$.
Here and elsewhere, a weight $\la$ will be identified by its Dynkin labels
$\la_1,\ldots,\la_n$. More generally, define $P_{++}(\T_{L,R})$ in the
obvious way for any type $\T$ (it will be the cartesian product of sets
in (2.8$b)$).

The affine Weyl group of $g\*{(1)}$ acts on $M\**$ as the semi-direct product
of the group of translations by vectors in $(k+h\*\vee)M$, with the (finite)
Weyl group of $g$. The affine Weyl orbit of any weight $\la$ intersects
$P_+(g,k+h\*\vee)$ in exactly one weight, call it $[\la]$. If $[\la]$ lies on
the boundary of $P_+$, define the parity $\eps(\la)=0$. Otherwise it lies in
the interior $P_{++}(g,k+\h)$, and there exists a unique affine Weyl
transformation $w$ such that $[\la]=w(\la)$; in this case define the parity
$\eps(\la)=det(w)=\pm 1$. Incidentally, the task of finding the values of
$[\la]$ and $\eps(\la)$ for arbitrary $\la,g,k$ is not difficult, even for
large ranks, and efficient algorithms exist for all algebras (see ref.~\RTW{}
for $A_n$).

Let $\T$ be any type satisfying eqs.~(2.2), (2.6) and (2.7$b$).
So $\T$ will in general include a null augment.
Define $M(\T)=M(\T_L)\oplus \sqrt{-1}M(\T_R)$,  an indefinite
lattice of dimension $n_L+n_R$; $x\in M(\T)$ can be written $(x_L;x_R)$
in the usual way. Consider any $\la_L\in P_{++}(\T_L), \,\la_R \in P_{++}
(\T_R)$. By $\la_L/\sqrt{k_L'}$ we mean the scaled vector
($\la_{L1}/\sqrt{k_{L1}'}, \ldots, \la_{Ll}/\sqrt{k_{Ll}'})$, using obvious
notation; similarly for $\la_R/\sqrt{k_R'}$.
Then $(\la_L/\sqrt{k_L'};$ $ \la_R/\sqrt{k_R'})\in M(\T)\**$. We call a
positive integer $L$ the order of $(\la_L; \la_R)$ in $\T$ when, for any
integer $m$, the vector $(m\la_L/\sqrt{k_L'};  m\la_R/\sqrt{k_R'})$ lies in
$M(\T)$ {\it iff\/} $L$ divides $m$. Let $N$ be any modular invariant of type
$\T$. Then for each  $\ell$ relatively prime to $L$,
$\eps_L(\ell\la_L)\eps_R(\ell\la_R)\ne 0$ and
$$N_{\la_L\la_R}=\eps_L(\ell\la_L)\eps_R(\ell\la_R)
N_{[\ell\la_L]_L[\ell\la_R]_R}.\eqno(2.9)$$

Eq.~(2.9) is called the parity rule for heterotic invariants. Its proof is
identical to the proof for non-heterotic invariants, given in ref.~\COMM,
which is based on the lattice method. (The one difference between the lattice
methods for heterotic and non-heterotic types which {\it seems} relevant is the
presence in the heterotic case of the translate $v$ -- see the following
section.  However, by augmenting $\T$ by an even number of $A_{1,0}$, we can
get $(2.7a)$ satisfied, in which case $v=0$ can always be chosen; this
augmenting will not affect $\eps_{L,R}$, and will affect $[-]_{L,R}$ only in a
trivial way.) For a different discussion of the heterotic parity rule, one
not involving augments, see ref.~\CG. There it is also generalized to any
RCFT.

The parity rule has two valuable consequences. All such
$([\ell\la_L]_L;\,[\ell\la_R]_R)$ form a family of essentially equivalent
representations, so that the coefficient $N_{\la_L\la_R}$ for just one
representative of each family need be stored.  Another important implication of
(2.9) is that for any $\ell$ coprime to $L$,
$$\eps_L(\ell\la_L)\eps_R(\ell\la_R)=-1\Rightarrow
N_{\la_L\la_R}=0\eqno(2.10a)$$ for any {\it physical\/} invariant $N$.

The parity rule simplifies the search for physical invariants by limiting the
modular invariants we need to consider. In particular, call $(\la_L;\la_R)$ a
{\it positive parity pair\/} if $\eps_L(\ell\la_L) =\eps_R(\ell\la_R)$ for all
$\ell$ coprime to the order $L$. By the {\it positive parity commutant\/} we
mean the subspace of the Weyl-folded commutant (the space of modular
invariants) consisting of all invariants $Z$ with the property that
$N_{\la_L\la_R}\ne 0$ only for positive parity pairs $(\la_L;\la_R)$. The
positive parity commutant contains all positive invariants and so is the only
part of the Weyl-folded commutant we need to consider. It is generally
significantly smaller than the full Weyl-folded commutant. We will need this
observation to simplify the analysis for some of the more complicated cases.

Now consider $\la_L=
\rho_L$, $\la_R=\rho_R$. Then if $\T$ is to be realized by a physical
invariant $N$, $(2.10a)$ and (P3) imply that
$$\forall\ell\sp{\rm coprime\sp to\sp}L,\sp \eps_L(\ell\rho_L)=\eps_R
(\ell\rho_R).\eqno(2.10b)$$
Eq.~(2.10$b)$ is a strong constraint on the candidates. In Table 1 we run
through all the candidates, and find the smallest positive $\ell$ violating
$(2.10b)$ (if one exists). This $\ell$ is listed in the Table.

The result is that there are precisely 14 non-AA-types which pass both
conditions (2.2) and (2.10$b$). Some of these have physical invariants,
some do not. In the following sections we will consider each of these in turn.
\bigskip \bigskip
	       \centerline{{\bf 3. The lattice method }}
\bigskip\nobreak
In this section we will first briefly review the extension of the
Roberts-\-Terao-\-Warner lattice method \WA{} to heterotic invariants. Their
method is a means of using self-dual lattices to generate invariants of the
form (1.1). It was originally designed for symmetric types, but has been
generalized \HET{} to heterotic ones using the idea of ``null augments''
discussed in subsect.~2.2. The method for symmetric types is summarized below
in eq.~(3.1), and for heterotic types in (3.2$b$). Next,  we apply it to
several of the candidates, finding not only all physical invariants for those
types, but also the entire commutant. The definitions of the few lattice
concepts we need can be found in ref.~\CS.

The Roberts-Terao-Warner \WA{} lattice method is a means of using self-dual
lattices to find modular invariants of the form (1.1). It was originally
designed for symmetric invariants, but has been generalized \HET{} to heterotic
invariants, using the idea of ``null augments'' discussed in subsect.~2.2.

We will begin by reviewing the symmetric case. For notational convenience
consider $g=g_L=g_R$ simple. Let $M$ be the coroot lattice of $g$.  Define the
indefinite lattice $\L_0=(\sqrt{k'}\,M)\oplus(\sqrt{-k'}\,M)$, where
$k'=k+h\*\vee$ is the height. Consider any even self-dual lattice
$\L\supset\L_0$, of equal dimension to $\L_0$. There will only be a finite
number of these $\L$. For each of them, there will only be a finite number of
cosets $[x]=[x_L;x_R] \in\L/\L_0$.  Choose any $x=(x_L;x_R)\in\L$, and put
$\la_L=\sqrt{k'}x_L$, $\la_R=\sqrt{k'}x_R$. Then $\la_{L,R}$ are weights of
$g$, \ie $\la_L,\la_R\in M\**$.

The Roberts-Terao-Warner method associates to every coset $[x]\in\L/\L_0$
the character product
$$
\eps(\la_L)\eps(\la_R)\chi_{[\la_L]}\chi_{[\la_R]}\**.\eqno(3.1)
$$
The partition function $WZ_{\L}$ corresponding to $\L$ consists of the sum of
terms $(3.1)$ over all cosets. Because $\L$ is even and self-dual, it is easy
to show $WZ_{\L}$ will be modular invariant. Indeed, it has been shown \COMM{}
that these $WZ_{\L}$ span the commutant of $g$, level $k$. Examples of
applications of this symmetric lattice method can be found in [12,13].

Unfortunately, extending this useful method to the heterotic types presents
certain complications. Again define $\L_0=(\sqrt{k_L'}\,M_L)\oplus
(\sqrt{-k_R'}\,M_R)$. In general, there will not exist any self-dual
lattices $\L\supset\L_0$. The necessary and sufficient condition for
that to happen is the {\it similarity} \Gal{} of the left and right
lattices: $\sqrt{k_L'}\,M_L\sim \sqrt{k_R'}\,M_R$.  Further, for a self-dual
indefinite lattice $\L$ to be {\it even} also, the condition (2.7$a$) must be
satisfied as well.

The way out of these complications is described in detail in \HET. We
will state here the conclusions. First, we must augment the heterotic type we
are interested in, by some level 0 algebras in such a way that the resulting
type satisfies (2.6). This is discussed in subsect.~2.2. Let the base lattice
$\L_0$ be defined with respect to this augmented type. The level 0 characters
are all identically equal to 1, so at the end of the calculation they do
not appear explicitly, but the null augments do affect the final lattice
partition function through the parities of their weights, and are necessary
to ensure modular invariance.

Eq.~(2.6$a$) guarantees we will have self-dual $\L\supset\L_0$, though
none of these may be even. Find a vector $v\in\L$ such that
$$x\*2+2x\cdot v\equiv 0\sp({\rm mod}\sp 2)\sp \forall x\in\L\eqno(3.2a)$$
(there are infinitely many such $v$, which one we choose will only affect
our final modular invariant by an irrelevant global sign). Then for
each coset $[x]\in\L/\L_0$ (again there will only be finitely many of these),
associate the character product
$$(-1)\*{x\*2}\eps(\la_L)\eps(\la_R)\chi_{[\la_L]_L}\chi_{[\la_R]_R}\**,
\eqno(3.2b)$$
where now $\la_{L}=\sqrt{k_L'}(x_L+v_L)$, and similarly for $\la_R$
(eq.~(3.2$b$) is meaningful because $\sqrt{k_L'}v_L$ is also a weight -- this
follows immediately from $(3.2a)$ and the fact that each coroot lattice $M$ is
even). The sum of all these terms $(3.2b)$, one for each coset, defines the
lattice partition function $WZ_{\L}\*v$. These are in fact modular invariant
\HET. It is proven in \HET{} that these $WZ_{\L}\*v$ again span the commutant
of the desired type. So in principle this reduces the problem of finding all
heterotic invariants of a given type to finding all self-dual $\L\supset\L_0$.
Examples of the heterotic lattice method can be found in \HET.

One remark should be made before we can proceed. Suppose $(2.7a)$ holds.  Then
there exist self-dual $\L\supset\L_0$ which are even. It can be shown \HET{}
that the $WZ_{\L}\*v$ associated with these even $\L$ also span the commutant.
Moreover,  by $(3.2b)$ we can choose $v=0$ for these $\L$.  We can always
fine-tune the ranks, by augmenting by $A_{1,0}$, so that $(2.7a$) is satisfied.
In some cases this can make the whole procedure a little more efficient.

The remainder of this section will be devoted to applying this lattice method
to candidates 1, 2, 3, 44 and 51. The task of finding all self-dual gluings
$\L$ of a given base lattice $\L_0$ is not very difficult, at least if $\L_0$
has a reasonably small dimension and determinant. The key is to exploit all the
(\eg{} Weyl) symmetries present.  We will discuss type 1 in detail, but we
first summarize the results for all these types. See also Table 2.

For type 1, we find that the commutant is spanned by the partition functions of
nine lattices ($9=3\cdot 3$, the second `3' reflects the fact that Aut($M(B_4))
/W(B_4)$ has order 3). These partition functions are not linearly independent,
and the commutant turns out to have dimension 3. For type 2, five lattices span
the commutant, which is only 2-dimensional. For type 3, there are $9\cdot
6\cdot 2$ lattices (the factor of 2 corresponds to an automorphism of the
lattice $\sqrt{6}{\bf Z}\*4$ of the augment $C_{2,0}C_{2,0}$); for type 44,
$6\cdot 2$ lattices; and for type 51, $5\cdot 2$ lattices.

We will work out the type 1 case explicitly; although it is a little simpler
than most of the others, it is complicated enough to include the features
of the general case.

The base lattice $\L_0$ for type 1 is $\L_0=\sqrt{21}D_4$, where $D_4$ is the
$D_4$ root lattice \CS, \ie the set of all even-normed vectors in $\Z\*4$ (no
null augment is needed in this case). The determinant $|\L_0|$ equals
$21\*4\cdot 4$, more precisely $\L_0\**/\L_0\cong \Z_4\times \Z_{21}\*4$, which
means that we need an order 2 vector $g_1\in \L_0\**$ and two independent order
21 vectors $g_2,g_3\in\L_0\**$; $\L$ will then be defined by
$$\L=\L_0[g_1,g_2,g_3]\equi \bigcup_{a=0}\*1 \bigcup_{b=0}\*{20}
\bigcup_{c=0}\*{20}
\{\L_0+ag_1+bg_2+cg_3\}.\eqno(3.3)$$
As long as $g_1,g_2,g_3$ are independent, \ie they generate (mod $\L_0$) a
group isomorphic to $\Z_2\times \Z_{21}\*2$, and have integer dot products with
each other, then $\L_0[g_1,g_2,g_3]$ will be self-dual.

There is an inaccessible number of triples $g_1,g_2,g_3$ with these properties,
but most of these yield identical partition functions $WZ\*v_{\L}$. Firstly, we
are only interested in triples that lead to different $\L$. Secondly, if two
triples $g_1,g_2,g_3$ and $g_1',g_2', g_3'$ give rise to self-dual lattices
$\L$, $\L'$ differing by a global Weyl-reflection, \ie $\exists w\in W(B_4)$
such that $w(\L)=\L'$, then by the Weyl-Kac character formula \KAC{} and
eq.(3.2$b$), $WZ\*v_\L$ and $Wz_{\L'}\*{v'}$ will differ by at most a global
sign.  We will find it convenient to ``modulo out'' Aut($D_4$), rather than its
subgroup $W(B_4)$, but at the end we must apply the three non-Weyl
automorphisms to each of the triples we have obtained.

First, let us list the possibilities (mod $\L_0$) for $g_1$: $\sqrt{21}(
{1\over 2},{1\over 2},{1\over 2}, {1\over 2})$, $\sqrt{21}(1,0,0,0)$,
and $\sqrt{21}({1\over 2},{1\over 2},{1\over 2},-{1\over 2})$ in the
standard orthonormal basis for $\Z\*4\supset D_4$. These are connected
by the triality of $D_4$, so (modulo Aut($D_4)$) there is a unique choice
for $g_1$: we may choose $g_1=\sqrt{21}(1,0,0,0)$. This will mean that
both $g_2$ and $g_3$ will lie in $1/\sqrt{21}\,\Z\*4$.

Our task to find $g_2$ and $g_3$ is simplified a bit by noting \CS{} that
there is only one self-dual (positive definite) lattice of dimension
4, namely $\Z\*4$. So what we must find are 4 orthonormal vectors
$u_i$ with coordinates $1/\sqrt{21}(a,b,c,d)$ (we will drop the $\sqrt{21}$
in the following). Up to signs and reorderings, there are only 2 different
unit vectors: (4,2,1,0) and (3,2,2,2).

Suppose first that there is at least one unit vector in $\L$ of the first
kind. Then up to $W(B_4$) we may write $u_1=(4,2,1,0)$. Up to irrelevant
sign differences, there are precisely 6 unit vectors orthogonal to $u_1$.
Running through these possibilities we find there are only 2 different
solutions: $u_2=(-2,4,0,1)$, $u_3=(0,-1,2,4)$ and $u_4=(-1,0,4,-2)$; and
$u_2=(0,-2,4,1)$, $u_3=(-2,3,2,-2)$ and $u_4=(-1,2,0,4)$. For both of these
solutions we may choose $g_2=u_1$, $g_3=u_2$ and $v=({1\over 2},{5\over 2},
{7\over 2},{3\over 2})$.

The remaining possibility is that all four $u_i$ are of the second kind
of unit vector. In this case there is (mod $W(B_4$)) only one $\L$, given
by $u_1=(3,2,2,2)$, $u_2=(-2,3,2,-2)$, $u_3=(-2,-2,3,2)$ and $u_4=(-2,2,-2,3)$.
Here we may choose $g_2=u_1$, $g_3=u_2$, and $v=({-3\over 2},{5\over 2},
{5\over 2},{5\over 2})$.

\bigskip\bigskip\centerline{\bf 4. Maximal chiral extensions and heterotic
invariants} \bigskip\nobreak
The lattice method is quite practical as long as the levels and ranks do not
get too large. But presumably it is completely inappropriate for candidates
like $n=59$ in Table 1. And it cannot be applied to an infinite family of
types, like for instance the AA-types.  In these cases, we need a more
theoretical approach. One such approach is suggested by the work of \MS.

The techniques discussed in this section are very powerful. However, they come
with two {\it caveats}. The main one is that they require a complete knowledge
of the physical invariants of the symmetric types $(\T_L;\T_L)$ and
$(\T_R;\T_R)$. The other is that additional conditions, beyond (P1)-(P3) of
sect.~1, must be imposed. These conditions are discussed in detail in
subsect.~4.1 below, and they all are physically valid, but there are reasons
for preferring classifications with a minimum of imposed conditions.

\bigskip \noindent{{\it 4.1 The cardinality test}}\medskip\nobreak
We will collectively call the techniques contained in this subsection the
``cardinality test'', even though only one actually involves comparing
cardinalities.

Let ${\cal C}\*{L},{\cal C}\*R$ denote the {\it maximally extended chiral
algebra\/} \MS{} of respectively the holomorphic, anti-holomorphic
sector of the theory. Let $ch_i$, $i=1,\ldots,a$ and
$\tilde{ch}_j$, $j=1,\ldots,b$, be the characters of ${\cal C}\*L$ and
${\cal C}\*R$, respectively. Label these so that $ch_1$ and $\tilde{ch}_1$
correspond to the identities.  $ch_i,\tilde{ch}_j$ can be expressed as linear
combinations, over the non-negative integers, of the affine characters
$\c_\la\*{(L)},\c_\mu\*{(R)}$ of $\T_L$ and $\T_R$, respectively:
$$
ch_i=\sum_{\la} m_{i\la}\chi_\la\*{(L)},\quad
\tilde{ch}_j=\sum_{\mu} \tilde{m}_{j\mu}\chi_\mu\*{(R)}.\eqno(4.1a)$$
We have $m_{i\rho_L}=\delta_{i1}$ and $\tilde{m}_{j\rho_R}=\delta_{j1}$.
Let the $S$ and $T$ modular matrices for these extended algebras be denoted
$S\*{(e)},T\*{(e)},
\tilde{S}\*{(e)},\tilde{T}\*{(e)}$. Then
$$S\*{(e)\dag}S\*{(e)}=I,\sp S\*{(e)T}=S\*{(e)}, \sp
T\*{(e)\dag}T\*{(e)}=I,\sp T\*{(e)T}=T\*{(e)}, \sp
S\*{(e)}_{1i}\geq S\*{(e)}_{11}>0,\eqno(4.1b)$$
with similar expressions for $\tilde{S}\*{(e)},\tilde{T}\*{(e)}$.

Finally, the partition function (1.1) of any physical theory will look like
$$Z=\sum_{i=1}\*a ch_i \tilde{ch}_{\sigma i}\**,\eqno(4.2)$$
for some bijection $\sigma$. This means that the numbers of characters in the
algebras ${\cal C}\*L$ and ${\cal C}\*R$   must be equal, $a=b$, and that
$$S\*{(e)}_{ij}=\tilde{S}\*{(e)}_{\sigma i,\sigma j},\quad
T\*{(e)}_{ij}=\tilde{T}\*{(e)}_{\sigma i,\sigma j},\eqno(4.3)$$
for all $1\leq h,i,j\leq a$.

So far in this paper all of our arguments have assumed only the familiar
properties (P1-P3). Ref.~\MS{} assumes a little more (namely duality, which
is required for any theory to be physical). To make clear precisely which set
of assumptions are being made, we will call an invariant {\it physical\/} if it
obeys (P1-P3), and {\it strongly physical\/} if in addition it obeys
(4.1),(4.2) (and hence (4.3)). There are physical invariants which are not
strongly physical, but  to be physically acceptable an invariant must be
strongly physical.

These are far-reaching facts. We will use them in the following way.
Suppose a given type $\T=(\T_L;\T_R)$ has a strongly physical invariant $Z$,
\ie maximal chiral extensions ${\cal C}\*L,{\cal C}\*R$ obeying (4.1),
(4.2) for some bijection $\sigma$. Construct the function
$$Z_L=\sum_{i=1}\*a |ch_i|\*2.\eqno(4.4a)$$
Then from the previous comments we know that $Z_L$ is a modular invariant
of type $(\T_L;\T_L)$. Similarly, we can construct a modular invariant
$Z_R$ of type $(\T_R;\T_R)$. Expanded in terms of affine characters,
$Z_L,Z_R$ are in block form:
$$Z_L=\sum_{i=1}\*a |\sum_{\la} m_{i\la}\chi_\la\*{(L)}|\*2=\sum_{\la,\la'}
M\*{(L)}_{\la\la'}\chi_\la\*{(L)}\chi_{\la'}\*{(L)*},\eqno(4.4b)$$
with a similar expression for $Z_R$. Note that $M\*{(L)}_{\la\la'}\geq 0$,
and $M_{\rho_L\rho_L}\*{(L)}=1$.

Suppose we know all strongly physical invariants of (non-heterotic) types
$(\T_L; \T_L)$ and $(\T_R; \T_R)$. Then in order  for there to be a strongly
physical heterotic invariant of type $(\T_L; \T_R)$, there must be
strongly physical invariants of types $(\T_L;\T_L)$ and $(\T_R;\T_R)$
with the same number of maximally extended characters. We will
call this the {\it cardinality test}.

For example, consider candidate $n=59$: $\T=(A_{2,105};G_{2,5})$. All
strongly physical invariants are known \GA{} for $(A_{2,105};A_{2,105})$,
and all physical invariants are known \HK{} for $(G_{2,5};G_{2,5})$.
In particular, there are precisely 4 strongly physical invariants for
$A_{2,105}$, two with $a=106\cdot 107/2=5671$ and two with $a=35\cdot
36/2+3=633$. There is precisely 1 physical invariant for $G_{2,5}$, and it has
$a=(7\cdot 8/2-4)/2=12$ extended characters. Thus there can be no strongly
physical invariants of type 59.

This analysis will allow us to handle the infinite series of AA-types:
$$\T=(\{A_{1},k_1\},\{A_1,k_2\}; \{A_1,k_3\},\{A_1,k_4\}).$$
Here, the central charge condition (2.2) becomes (2.4). Define $s=k_1'+k_2'$,
$p=k_1'k_2'$, $s'=k_3'+k_4'$ and $p'=k_3'k_4'$, then $s,p$ uniquely specify
$k_1',k_2'$, up to order; in particular they are the 2 roots of
$k'{}\*2-sk'+p=0$. (2.4) can be rewritten as $$s'=p's/p.\eqno(4.5)$$

The $A_1+A_1$ (non-heterotic) physical invariants have been classified in
ref.~\SUSU, together with some anomolous levels worked out in ref.~\HK{} and
here in the appendix. The result is that there are a number of exceptionals,
which we will discuss later, along with the simple current invariants and their
conjugations. Let us consider first the (maximally extended) chiral algebras of
the simple current invariants (conjugations, being automorphisms, do not affect
the chiral algebra). There are 5 of these: ${\cal C}\*{(1)}_{k\ell}=A_{1,k}
+A_{1,\ell}$, the unextended chiral algebra, defined for all levels $k,\ell$;
${\cal C}\*{(2)}_{k\ell}$, defined for $k\equiv -\ell\equiv \pm 1$ (mod 4), is
a chiral algebra corresponding to the simple current $J=(1,1)$; ${\cal
C}\*{(3)}_{k\ell}$, defined for $k\equiv \ell\equiv 0$ or 2 (mod 4),
corresponding to $J=(1,1)$; ${\cal C}\*{(4)}_{k\ell}$, defined for $k\equiv 0$
(mod 4), corresponding to $J=(1,0)$; and finally ${\cal C}\*{(5)}_{k\ell}$,
defined for $k\equiv \ell\equiv 0$ (mod 4), corresponding to $J=(1,0)$ and
$J'=(0,1)$. These have cardinalities given by:
$$\eqalignno{ {\rm
card}\,\,{\cal C}\*{(1)}_{k\ell}:&\sp (k+1)(\ell+1);&(4.6a)\cr {\rm
card}\,\,{\cal C}\*{(2)}_{k\ell}:&\sp {(k+1)(\ell+1)\over 4};&(4.6b)\cr {\rm
card}\,\,{\cal C}\*{(3)}_{k\ell}:&\sp {k\ell+k+\ell+8\over 4};&(4.6c)\cr {\rm
card}\,\,{\cal C}\*{(4)}_{k\ell}:&\sp {(k+8)(\ell+1)\over 4};&(4.6d)\cr {\rm
card}\,\,{\cal C}\*{(5)}_{k\ell}:&\sp {(k+8)(\ell+8)\over 16}.&(4.6e)
\cr}$$
We will show that there cannot be a heterotic invariant with chiral algebras
${\cal C}\*L={\cal C}\*{(\alpha)}_{k_1k_2}$, ${\cal C}\*R=
{\cal C}\*{(\beta)}_{k_3k_4}$,
for any $\alpha,\beta=1,\ldots,5$. These cardinalities alone suffice to handle
some cases. For
example, consider $\alpha=\beta=1$. Then the cardinality condition becomes
$(k_1+1)
(k_2+1)=(k_3+1)(k_4+1)$, \ie $p-s+1=p'-s'+1$, which combined with (4.5)
forces $p=p'$, $s=s'$, in other words the sets $\{k_1,k_2\}$ and $\{k_3,k_4\}$
are equal. The arguments
handling $\alpha=\beta=2$, $\alpha=\beta=3$ and $\alpha=\beta=5$ are identical.

But using cardinalities alone is too difficult in some cases. Fortunately, for
all simple current extensions it is trivial to compute their $S$-matrix
elements $S\*{(e)}_{1i}$:
$$S\*{(e)}_{1,i}={\|{\cal J}\|\over F_{i}}S_{\rho,\la_i},\eqno(4.7)$$
where $S$ is the $S$-matrix for the underlying affine algebra,
$\la_i$ is any affine weight satisfying $m_{i\la}\ne
0$ (see (4.1$a$)), $\|{\cal J}\|$ is the number of simple currents in the
theory, and $F_i$ is the number of these simple currents fixing $\la_i$.

{}From this calculation we can read off the number of solutions $i$ to the
equation
$S\*{(\alpha)}_{1,i}=S\*{(\alpha)}_{1,1}$ ($S\*{(\alpha)}$ here is the
$S$-matrix for the simple current extension ${\cal C}\*{(\alpha)}_{k\ell}$):
there are 4 solutions for $\alpha=1$ (namely, those $i$ with $\la_i=(1,1),
(k+1,1),(1,\ell+1)$, and
$(k+1,\ell+1)$); there is only 1 solution for $\alpha=2$
(namely, $\la_i=(1,1)$); there are 2 solutions for $\alpha=3$ ($\la_i=(1,1)$
and $\la_i=(k+1,1)$);
there are 2 solutions for $\alpha=4$ ($\la_i=(1,1)$
and $\la_i=(1,\ell+1)$); and there is only one solution for
$\alpha=5$ ($\la_i=(1,1)$).
There are some low level exceptions, due to fixed points, to these numbers:
${\cal C}\*{(3)}_{22}$ has 4 solutions; ${\cal C}_{4,\ell}\*{(4)}$ has 4; and
${\cal C}\*{(5)}_{4,\ell}$ has 3, unless $\ell=4$ in which case it has 9.
These exceptions can be handled separately, \eg by explicitly solving (2.4)
(there are precisely 53 different heterotic solutions to (2.4) with $k=4$).

By $(4.3$), this leaves only the following possibilities:

\item{(i)} ${\cal C}\*L={\cal C}\*{(2)}$, ${\cal C}\*R={\cal C}\*{(5)}$;
\item{(ii)} ${\cal C}\*L={\cal C}\*{(4)}$, ${\cal C}\*R={\cal C}\*{(4)}$;
\item{(iii)} ${\cal C}\*L={\cal C}\*{(3)}$, ${\cal C}\*R={\cal C}\*{(4)}$.

Consider possibility (ii) (the arguments for (i) and (iii) are similar).
We may assume $k_1,k_3>4$. The {\it second} smallest value of
$S\*{(4)}_{1,i}$ will be realized by $i=i'$, where $\la_{i'}=(1,2)$
(unless $\ell=1$, or
sometimes when $k=8$). Then for $k_2,k_4>1$ and $k_1,k_3>8$, dividing
$S\*{(4)}_{1,i'}
=\tilde{S}\*{(4)}_{1,{\tilde i}'}$ by $S\*{(4)}_{1,1}=
\tilde{S}\*{(4)}_{1,1}$, and using (4.7), gives $\sin(2\pi/k_2')/
\sin(\pi/k_2')=\sin(2\pi/k_4')/\sin(\pi/k_4')$, \ie $k_2'=k_4'$; (2.4) then
forces $(k_1,k_2)=(k_3,k_4)$. If however $k_2=1$, then (2.4) can be solved
explicitly: there are precisely 9 heterotic solutions; of these, only one
satisfies the congruences $k_1\equiv k_3\equiv 0$ (mod 4), and that one fails
the cardinality test (4.6). $k_1=8$ succumbs to a similar argument (there
are precisely 110 solutions to (2.4) with $k=8$).

Thus there can be no (strongly) physical invariants whose extended chiral
algebras are both simple current extensions. This leaves the exceptional
extensions. The greatest source of these involves the ${\cal E}_{10}$ or ${\cal
E}_{28}$ exceptionals \CIZ{} of $A_1$ (we can avoid ${\cal E}_{16}$ because it
corresponds to a simple current extension).  It is not difficult, using
eqs.~(2.3), to run through on a computer all solutions to (2.4) with $k_1=10$
or $k_2=28$ (there are 183 and 676 different heterotic solutions for these,
respectively), and then explicitly check the cardinality test for each solution
(the cardinality of the ${\cal E}_{10}$ extension is 3, and that of ${\cal
E}_{28}$ is 2; when $k_1=10$ and $k_2
\equiv 2$ (mod 4) there is another cardinality, namely $(3k_2+10)/4$
corresponding to $({\cal E}_{10}\otimes{\cal A}_{k_2})\,N_{10,k_2}(J_1J_2)$,
which also must be considered). In this way, we find that
there is only one AA-type made up of any combination of ${\cal E}_{10}$,
${\cal E}_{28}$ and the simple currents that passes the cardinality test. It
is candidate 62 in Table 1.

The remaining (non-heterotic) exceptionals for $A_1+A_1$ (see \GHF{}
and the appendix below) occur at levels $(k_1,k_2)=(4,4)$, (6,6), (8,8),
(10,10), (2,10), (3,8), (3,28) and (8,28). We do not have to consider the
(4,4), (8,8), (2,10) or (3,28) exceptionals since they correspond to
automorphisms of chiral algebras already considered. The cardinalities of the
remaining exceptional chiral algebras can be read off, and are: 3,
4, 4, 2 respectively. It is an easy task to find the solutions to
(2.4) for each of these levels, and to explicitly compare cardinalities.
None of these pass the cardinality test.

\bigskip\noindent{{\it 4.2 Heterotic vrs symmetric automorphisms}}\medskip
\nobreak
Eqs.~(4.1), (4.2), and (4.3) can have more to say, even when the candidate
passes the cardinality test. In particular, suppose we have found a heterotic
invariant (4.2), corresponding to chiral extensions ${\cal C}\*L$ of
$\T_L$ and ${\cal C}\*R$ of $\T_R$. Then by (4.3) the modular properties
of ${\cal C}\*L$ and ${\cal C}\*R$ are completely equivalent -- the bijection
$\sigma$ makes this equivalence explicit.

Let us now ask the question: how many other heterotic invariants are there
with the given chiral extensions? In other words, how many other bijections
$\sigma'$ can we find? The answer is completely known, if we know all
(non-heterotic) strongly physical invariants of $\T_L$, say. Let $Z_1,\ldots,
Z_n$ be those physical invariants of $\T_L$ with (maximal) LHS and RHS
chiral algebra ${\cal C}\*L$. Each of them is given by a bijection $\sigma_i$,
mapping the characters of ${\cal C}\*L$ to themselves (\eg{} one of these
bijections, the one corresponding to the physical invariant of ($4.4a$),
will be the identity). Then there will be precisely $n$ heterotic invariants
with chiral algebras ${\cal C}\*L$ and ${\cal C}\*R$, and they will be
given by the bijections $\sigma_i\circ \sigma$.

The only potential problem with this idea is field identification -- \ie{} in
some cases different chiral characters $ch_i\ne ch_j$ correspond via $(4.1a$)
to identical expressions of affine characters. But in practice this rarely
presents any complications (see the example below).  In this way we can write
down all heterotic strongly physical invariants for candidates 39, 57 and 60.
These are given in eqs.~(6.1$d)$, (6.1$f$), $(6.1g)$.

We will give one example here. Consider $n=39$. There are conformal embeddings
\CE{} $C_{2,3}\subset D_{5,1}$ and $A_{1,10}A_{1,10}\subset D_{5,1}$, so each
physical invariant of $(D_{5,1};D_{5,1}$) will yield a physical invariant of
$(C_{2,3};A_{1,10}A_{1,10})$. There are only two \COMM{} physical invariants of
$(D_{5,1};D_{5,1})$, and they both correspond to the heterotic invariant given
in $(6.1d)$. We know \HK{} all the physical invariants of $C_{2,3}$ and
$A_{1,10}A_{1,10}$, in particular each has only one corresponding to this
chiral extension.

There is field identification present here, but all the bijections map the
identified fields to each other, so there is only one heterotic invariant.

Incidentally, both $C_{2,3}$ and $A_{1,10}A_{1,10}$ also have physical
invariants with chiral cardinalities of 10 (see eqs.~(4.3$j$) and $(4.4a$) in
\HK). However, there can be no bijection between them, and hence no
corresponding heterotic invariant. One way to see this is (4.3), the condition
$T\*{(e)}_{ij}=\tilde{T}\*{(e)}_{\sigma i,\sigma j}$ (see (5.1) below).

\bigskip\bigskip\centerline{\bf 5. The remaining candidates}
\bigskip\nobreak
In this section we discuss two more techniques, which suffice to complete
the classification of the rank 4 heterotic invariants.

\bigskip\noindent{{\it 5.1 Explicit calculations}}\medskip
For heterotic type (2.1), $T$-invariance becomes the selection rule
$$N_{\la\mu}\ne 0\Rightarrow \sum_{i=1}\*l {\la\*2_i\over k_{Li}'}-
\sum_{j=1}\*r{\mu\*2_j\over k'_{Rj}}\equiv \sum_{i=1}\*l{\rho\*2_{Li}\over
h_{Li}\*\vee}-\sum_{j=1}\*r{\rho\*2_{Rj}\over h\*\vee_{Rj}}\sp({\rm mod}\,\,
2).\eqno(5.1)$$

Some candidates have small enough levels so that they can be explicitly worked
out by hand. For this purpose the parity rule can also come in handy,
simplifying the work by reducing the numbers of independent variables. The idea
is to first use $T$-invariance to find all possible combinations
$\c_{\la}\*{(L)}\c_{\mu}\*{(R)*}$; the parity rule (2.9) can then find which of
these terms can be ignored, and which of the remaining coefficients
$N_{\la\mu}$ are independent. If this number is sufficiently small,
$S$-invariance can be done explicitly. Outer automorphisms (\ie simple
currents) can also be used to good effect here.

For example, consider candidate $n=45$. $T$-invariance tells us the modular
invariant will look like
$$Z=a\c_{11}\c_1\**\c_1\**+b\c_{11}\c_2\**\c_4\**+c\c_{12}\c_1\**\c_3\**
+d\c_{12}\c_2\**\c_2\**,\eqno(5.2a)$$ where the three characters in each term
correspond to $G_{2,1}$, $A_{1,1}$ and $A_{1,3}$ respectively. The parity rule
tells us that $$a=c,\sp b=d.\eqno(5.2b)$$ Thus we have only two independent
variables, $a$ and $b$. Note from $(5.2a$) that in each term the weight labels
for $A_{1,1}$ and $A_{1,3}$ are always equivalent to each other mod 2. From the
relation $N=S_LNS\*{\dag}_R$ we then get that $N_{\la;bc}=N_{\la;3-b,5-c}$, \ie
$a=d$ and $b=c$.  $S$-invariance can be explicitly checked to verify that the
resulting partition function is indeed modular invariant.

Hence for this candidate the commutant is one-dimensional, given by
the $Z$ in (5.2$a$) with $a=b=c=d$, and there is exactly one physical
invariant. It is listed in $(6.1e)$.

Similar arguments (\ie using (5.1), the parity rule, outer automorphisms,
and as a last resort explicit $S$-matrix calculations)
work for candidates 11, 12, 13 and 62.

\bigskip\noindent{{\it 5.2 Projection}}\medskip
Let $N$ be any physical invariant of type $(G_{2,24}G_{2,24};0)$.
Then by the usual projection argument \BOW,
$$M_{\la\mu}=\sum_{\nu}N_{\la\nu}N_{\nu\mu},\quad
M_{\la\mu}'=\sum_{\nu}N_{\la\nu}N_{\mu\nu}\eqno(5.3a)$$
will both be modular invariants of type $(G_{2,24};G_{2,24})$. In
fact, they will be {\it positive} invariants, \ie each $M_{\la\mu},M_{\la\mu}'
\ge 0$, and will be nonzero because $M_{\rho\rho},M_{\rho\rho}'
\geq N_{\rho\rho}=1$.

In Table 1 of \HK{} we find that the {\it positive parity
commutant}, which of necessity contains all positive invariants, is
one-dimensional for $g_L=g_R=G_{2,24}$ (indeed, for all $G_{2,k}$,
$5\le k\le 31$). Thus we get
$$M_{\la\mu}=a\delta_{\la\mu},\quad M'_{\la\mu}=a'\delta_{\la\mu}\quad
\forall \la,\mu\eqno(5.3b)$$
for some constants $a,a'>0$. Hence
$$\eqalignno{a=\sum_{\nu}N_{\la\nu}N_{\nu\la},\quad &a'=\sum_{\nu}
N_{\la\nu}\*2,\quad \forall \la&(5.4a)\cr
0=\sum_{\nu}N_{\la\nu}N_{\nu\mu}=&\sum_{\nu}N_{\la\nu}N_{\mu\nu},
\quad \forall \la\ne\mu.&(5.4b)\cr}$$
Therefore, $N_{\la\mu}\ne 0$ implies $N_{\nu\mu}=N_{\mu\nu}=0$ $\forall \nu
\ne \la$. From $(5.4a)$ this forces $a=a'=1$, so each $N_{\la\mu}=0$ or 1,
and there is only one 1 on each row and column of $N$.
In other words, there exists a permutation $\sigma$ of $P_{++}(G_{2,24})$
such that
$$N_{\la\mu}=\delta_{\mu,\sigma\la}, \forall\la,\mu.\eqno(5.5)$$

Let us investigate now the commutation of $N$ with $T$. Let $(m,n)$ be the
Dynkin labels of $\la$, and $(m',n')$ those of $\sigma\la$. Then $T$-invariance
(5.1) means $(\la\*2+(\sigma\la)\*2)/28\equiv 1/3$ (mod 2), \ie
$$m\*2+mn+n\*2/3+m'{}\*2+m'n'+n'{}\*2/3\equiv {14\over 3} \sp({\rm mod}\sp 28).
\eqno(5.6)$$
Choose $\la=(1,3)$, then (5.6) says among other things that $n'{}\*2\equiv
2$ (mod 3), which has no solution.

Thus $T$-invariance for (5.5) cannot be satisfied, so there are no physical
invariants for candidate 9.

\bigskip\bigskip\centerline{\bf 6. Conclusion}
\bigskip\nobreak
In this paper, we have developed several general techniques for classifying
heterotic physical invariants, and have applied them to find all such
invariants of total
rank 4. In addition, we have also determined all non-heterotic physical
invariants for $(A_{1,k_1}+A_{1,k_2};$ $A_{1,k_1}+A_{1,k_2})$ with $k_1\neq k_2
<22$ and $k_1=28$, $k_2<22$; these results are given in the Appendix.

The following is the list of all total rank 4 heterotic (strongly) physical
invariants:

$$\eqalignno
{(C_{4,10};-):\sp Z_2=&\c_{1111}+\c_{1,1,11,1}+\c_{1135}+\c_{1151}+
\c_{1155}+\c_{5113}+\c_{5117}&\cr
&+\c_{1,1,11,1}+\c_{1371}+\c_{1432}+\c_{1434}+\c_{1611}+\c_{5251}+\c_{5252}&\cr
&+\c_{1616}+\c_{1911}+\c_{1913}+\c_{2162}+\c_{2541}+\c_{5331}+\c_{6124}&\cr
&+\c_{2542}+\c_{3115}+2\c_{3333}+\c_{3353}+\c_{3413}+\c_{7313}&\cr
&+\c_{3415}+\c_{4142}+\c_{4144}+\c_{4522}+\c_{4523}+\c_{11,1,1,1};&(6.1a)\cr
(A_{2,3};A_{1,2}A_{1,10}):\sp Z_{11}=&(\c_{11}+\c_{14}+\c_{41})\{\c_1\c_1+
\c_1\c_7+\c_3\c_5+\c_3\c_{11}\}\**&\cr
&+\c_{22}\{\c_1\c_5+\c_1\c_{11}+\c_3\c_1+\c_3\c_7+2\c_2\c_4+2\c_2\c_8\}\**;
&(6.1b)\cr
(A_{2,4};A_{1,4},A_{1,12}):\sp Z_{13}=&\c_{11}(\c_1+\c_5)\**(\c_1+\c_{13})\**
+\c_{22}(\c_1+\c_5)\**(\c_5+\c_9)\**&\cr
&+(\c_{12}+\c_{21})\c_3\**\c_7\**+(\c_{24}+\c_{42})\c_3\**\c_7\**+
(\c_{14}+\c_{41})(\c_1+\c_5)\**\c_7\**&\cr
&+(\c_{13}+\c_{31})\c_3\**(\c_3+\c_{11})\**
+\c_{33}(\c_1+\c_5)\**(\c_3+\c_{11})\**&\cr
&+(\c_{15}+\c_{51})\c_3\**(\c_1+
\c_{13})\**+(\c_{23}+\c_{32})\c_3\**(\c_5+\c_9)\**;&(6.1c)\cr
(C_{2,3};A_{1,10}A_{1,10}):\sp Z_{39}=&(\c_{11}+\c_{32})\{\c_{1}\c_1+
\c_{1}\c_7+\c_{5}\c_5+\c_{5}\c_{11}&\cr
&\qquad\qquad\quad +\c_{7}\c_1+\c_{7}\c_7+\c_{11}\c_5+\c_{11}\c_{11}\}\**&\cr
&+(\c_{14}+\c_{31})\{\c_{1}\c_{11}+\c_1\c_5+\c_7\c_{11}+\c_7\c_5&\cr
&\qquad\qquad\quad +\c_5\c_7+\c_5\c_1+\c_{11}\c_1+\c_{11}\c_7\}\**&\cr
&+2\c_{22}\{\c_4\c_4+\c_4\c_8+\c_8\c_4+\c_8\c_8\}\**;&(6.1d)\cr
(G_{2,1};A_{1,1}A_{1,3}):\sp Z_{45}=&\c_{11}\{\c_1\c_1+\c_2\c_4\}\**+\c_{12}
\{\c_1\c_3+\c_2\c_2\}\**;&(6.1e)\cr
(A_{2,9};G_{2,3}):\sp Z_{57}=&(\c_{11}+\c_{1,10}+\c_{10,1}+\c_{55}
+\c_{52}+\c_{25})\{\c_{11}+\c_{22}\}\**&\cr&
+2(\c_{33}+\c_{36}+\c_{63})\cdot\c_{13}\**;&(6.1f)\cr
(C_{2,7};G_{2,4}):\sp Z_{60}=&(\c_{11}+\c_{16}+\c_{33}+\c_{72})
\{\c_{11}+\c_{14}\}\**+2(\c_{42}+\c_{44})\c_{22}\**&\cr
&+(\c_{13}+\c_{18}+\c_{34}+\c_{71})\{\c_{21}+\c_{15}\}\**.&(6.1g)\cr}
$$
The subscript `$n$' of $Z_n$ in these equations denotes the candidate number
(see Table 1); the subscripts on the $\c$'s denote the Dynkin labels.  The
invariant $Z_{2}$ was first found in ref.~\SCH, while $Z_{39}$, $Z_{45}$,
$Z_{57}$ and $Z_{60}$ are due to conformal embeddings \CE{} applied to the
diagonal invariants of $(D_{5,1};D_{5,1})$, $(G_{2,1};G_{2,1})$,
$(E_{6,1};E_{6,1})$ and $(D_{7,1};D_{7,1})$, respectively ($Z_{45}$ can also be
obtained from the rank 3 invariants (1.2$c$), (1.2$d$)). Both $Z_{11}$ and
$Z_{13}$ seem to be new.  $Z_{11}$ can be understood within the context of
\MS{} as a bijection between the simple current chiral extension of $A_{2,3}$,
and the chiral extension of $A_{1,2}A_{1,10}$ associated with the physical
invariant $({\cal A}_2\otimes {\cal E}_{10})\,N_{(2,10)}(J_1J_2)$; as such it
is intimately connected with the exceptional ${\cal E}''_{2,10}$ given in
$(A.6)$ below.  However, $Z_{13}$ is much harder to understand.

Our results demonstrate the scarcity of heterotic invariants. For example,
there are between 1 and 27 {\it non-heterotic} physical invariants
corresponding to each choice $(k_1,k_2)$ of level, for the algebra $A_1+A_1$;
however, there are zero {\it heterotic} physical invariants for
$g_L=g_R=A_1+A_1$, for any level. For total rank $\ge 5$, there will be
infinitely many heterotic invariants (just tensor non-heterotic ones with rank
3 or 4 heterotics), but their numbers will always be very small compared to the
non-heterotics of similar rank. This is reflected in the severity of the
constraints which must be satisfied by the algebras and levels of heterotic
physical invariants (see \eg (2.2) and (2.10$b$)); these constraints will be
trivially satisfied for non-heterotic invariants.

\bigskip\bigskip
\nobreak
This work was supported in part by the Natural Sciences and Engineering
Research Council of Canada. T.G.{} would like to thank the hospitality of the
IHES, as well as Antoine Coste, A.~N. Schellekens and Mark Walton for valuable
communications.
\vfill\eject
\def\d{\delta} \def\sp{\,\,\,}\def\i{\item}\def\o{\otimes}
\def\AA{{\cal A}}\def\DD{{\cal D}}\def\EE{{\cal E}}
\def\c{\chi}
\centerline{\bf Appendix}
\bigskip
In this Appendix we present the results of a computer search for the
non-heterotic physical invariants of the algebra  $A_{1,k_1}+A_{1,k_2}$ for
$k_1\neq k_2 <22$ and $k_1=28$, $k_2<22$. We used a method based on the notion
of even self-dual lattices developed in ref.~\HK{} (where, however, the
$A_1+A_1$ physical invariants had been calculated only at levels $k_1=k_2<22$).

Most of the physical invariants of this algebra belong to one of the infinite
series of invariants, some of which are obtained as tensor products of the
${\AA_k}$ and ${\DD_k}$ physical invariants of $A_{1,k}$ \CIZ.  But a number of
others not belonging to these series also occur at various levels. Together
with the four exceptionals given in eqs.~(4.3$f$)-$(4.3i)$ of \HK, and the
conjugations of all these invariants if $k_1=k_2$, the following list exhausts
all $A_1+A_1$ (non-heterotic) physical invariants, for all levels $k_1,k_2$.

We list first the infinite series of $A_1+A_1$ invariants. If the levels $k_1$
and $k_2$ are both even, there are 6 series; otherwise, only 2.  Let
$p_1=k_1+2$, $p_2=k_2+2$, $k=(k_1,k_2)$.  The invariants can be defined by
their corresponding coefficient matrices $N_{ij,i'j'}$, where $(ij)$ and
$(i'j')$ are the Dynkin labels of  the 2 weights $\lambda_L$ and $\lambda_R$,
with $0<i,i'<p_1$ and $0<j,j'<p_2$. We will name the invariants using
simple current notation. By $J_1$ we mean the simple current $(1,0)$, etc.

\medskip For $k_1\equiv k_2$ (mod 4), both odd, the 2 series are:
\i{(i)} the diagonal (identity) invariant $N_k(0)={\cal A}_{k_1}\otimes {\cal
A}_{k_2}$;
\i{(ii)} $N_k(J_1J_2)$ given by
  $$  N_k(J_1J_2)_{ij,i'j'}=\left\{
\matrix{ \d_{ii'}\d_{jj'} & {\rm if}\sp i\equiv j \sp ({\rm mod}\sp 2)\cr
\d_{i,p_1-i'}\d_{j,p_2-j'} & {\rm otherwise}\cr} \right..\eqno(A.1)$$

For $k_1\not\equiv k_2$ (mod 4), both odd, we have
\i{(i)} the diagonal invariant $N_k(0)$;
\i{(ii)} the invariant $N_k(J_1J_2)$ defined by
    $$  N_k(J_1J_2)_{ij,i'j'}=\left\{ \matrix{
                0 & {\rm if}\sp i\not\equiv j \sp
                  ({\rm mod}\sp 2)\cr
             \d_{ii'}\d_{jj'}+\d_{i,p_1-i'}\d_{j,p_2-j'} & {\rm otherwise}
                                      \cr}\right. .\eqno(A.2)$$

For $k_1$ even, $k_2$ odd, the 2 series (these are equal if $k_1=2$) are
\i{(i)} the diagonal invariant $N_k(0)$;
\i{(ii)} $N_k(J_1)={\cal D}_{k_1}\otimes {\cal A}_{k_2}$.

\medskip For $k_1\equiv k_2\equiv 2$ (mod 4), there are 6 series (3 if $k_1$
or $k_2$, but not both, equal 2; 2 if both $k_1=k_2=2$):
\i{(i)} the diagonal invariant $N_k(0)$;
\i{(ii)} $N_k(J_1)={\DD}_{k_1}\otimes {\AA}_{k_2}$;
\i{(iii)}  $N_k(J_2)=\AA_{k_1}\otimes \DD_{k_2}$;
\i{(iv)}  $N_k(J_1J_2)$ defined by eq.~$(A.2)$;
\i{(v)}  $N_k(J_1;J_2)=\DD_{k_1}\otimes \DD_{k_2}$;
\i{(vi)}  $N_k(J_1;J_1J_2)$ defined by
$$        N_k(J_1;J_1J_2)_{ij,i'j'}=\left\{ \matrix{
              0 & {\rm if}\sp i\not\equiv j\sp({\rm mod}\sp 2)\cr
             \d_{ii'}\d_{jj'}+\d_{i,p_1-i'}\d_{j,p_2-j'} &
                     {\rm if}\sp i\equiv j\equiv 1\sp({\rm mod}\sp 2)\cr
              \d_{i,p_1-i'}\d_{jj'}+\d_{ii'}\d_{j,p_2-j'} &
                     {\rm if}\sp i\equiv j\equiv 0 \sp({\rm mod}\sp 2)\cr}
                             \right.. \eqno(A.3)$$

For $k_1\equiv k_2\equiv 0$ (mod 4), the 6 series are:
\i{(i)} the diagonal invariant $N_k(0)$;
\i{(ii)}  $N_k(J_1)=\DD_{k_1}\otimes \AA_{k_2}$;
\i{(iii)}  $N_k(J_2)=\AA_{k_1}\otimes \DD_{k_2}$;
\i{(iv)}  $N_k(J_1J_2)$ defined by eq.~($A.2$);
\i{(v)}  $N_k(J_1;J_2)=\DD_{k_1}\otimes \DD_{k_2}$;
\i{(vi)} $N_k(aut)$ defined by
$$        N_k(aut)_{ij,i'j'}=\left\{ \matrix{
              \d_{ii'}\d_{jj'} & {\rm if}\sp i\equiv j\equiv 1 \sp
                      ({\rm mod}\sp 2)\cr
              \d_{i,p_1-i'}\d_{j,j'} & {\rm if}\sp i\not\equiv j\equiv 0\sp
                      ({\rm mod}\sp 2)\cr
              \d_{ii'}\d_{j,p_2-j'} & {\rm if}\sp i\not\equiv j\equiv 1\sp
                       ({\rm mod}\sp 2)\cr
              \d_{i,p_1-i'}\d_{j,p_2-j'} & {\rm if}\sp i\equiv j\equiv 0\sp
                       ({\rm mod}\sp 2)\cr}\right..\eqno(A.4)$$

For $k_1\equiv 0$, $k_2\equiv 2$ (mod 4), the 6 series (3 if $k_2=2$) are:
\i{(i)} the diagonal invariant $N_k(0)$;
\i{(ii)} $N_k(J_1)=\DD_{k_1}\otimes \AA_{k_2}$;
\i{(iii)} $N_k(J_2)=\AA_{k_1}\otimes \DD_{k_2}$;
\i{(iv)} $N_k(J_1J_2)$ defined by eq.~($A.1)$;
\i{(v)} $N_k(J_1;J_2)=\DD_{k_1}\otimes \DD_{k_2}$;
\i{(vi)} the invariant $N_k(J_2;J_1J_2)$ defined by
$$          N_k(J_2;J_1J_2)_{ij,i'j'}=\left\{ \matrix{
                \d_{ii'}\d_{jj'} & {\rm if}\sp i\equiv j\equiv 1\sp
                ({\rm mod}\sp 2)\cr
                \d_{ii'}\d_{j,p_2-j'}& {\rm if}\sp i\equiv j\equiv 0\sp
                ({\rm mod}\sp 2)\cr
                \d_{i,p_1-i'}\d_{j,p_2-j'} & {\rm if}\sp i\not\equiv j
                \equiv 1\sp({\rm mod}\sp 2)\cr
                \d_{i,p_1-i'}\d_{jj'} & {\rm if}\sp i\not\equiv j\equiv 0\sp
                ({\rm mod}\sp 2)\cr}\right.. \eqno(A.5) $$

Besides these simple current invariants, we also obtain solutions built on the
three $A_1$ exceptional invariants $\EE_{10}$, $\EE_{16}$, and
$\EE_{28}$. Namely, the three isolate solutions
$\EE_{10}\o\EE_{16}$, $\EE_{10}\o\EE_{28}$, and $\EE_{16}\o\EE_{28}$, and
the following  series for all $k_2$: $\EE_{10}\o\AA_{k_2}$,
$\EE_{16}\o\AA_{k_2}$, and $\EE_{28}\o\AA_{k_2}$; and for even $k_2\geq 4$:
$\EE_{10}\o\DD_{k_2}$, $\EE_{16}\o\DD_{k_2}$, $\EE_{28}\o\DD_{k_2}$, and
$(\EE_{10}\o\AA_{k_2}) \,N_{(10,k_2)}(J_1J_2)$.

Finally, there exist sporadic exceptional invariants not of the above types,
which we denote by $\EE''_{k_1,k_2}$. They are:
$$\eqalignno{
 \EE''_{2,10}\equi& |\c_{1,1}+\c_{1,7}+\c_{3,5}+\c_{3,11}|\*2
              +(\c_{1,5}+\c_{1,11}+\c_{3,1}+\c_{3,7})(\c_{2,4}+\c_{2,8})\**\cr
              &+ (\c_{2,4}+\c_{2,8})(\c_{1,5}+\c_{1,11}+\c_{3,1}+\c_{3,7})\**
              +|\c_{2,4}+\c_{2,8}|\*2 \;; & (A.6)\cr
\EE''_{3,8}\equi& |\c_{1,1}+\c_{3,5}+\c_{1,9}|\*2+
            |\c_{2,5}+\c_{4,1}+\c_{4,9}|\*2+|\c_{1,5}+\c_{3,3}+\c_{3,7}|\*2\cr
           &+|\c_{2,3}+\c_{4,5}+\c_{2,7}|\*2 \;; & (A.7)\cr
\EE''_{3,28}\equi& |\c_{1,1 }+\c_{1,11}+\c_{1,19}+\c_{1,29}|\*2
    +|\c_{2,7}+\c_{2,13}+\c_{2,17}+\c_{2,23}|\*2\cr
    &+ |\c_{3,7}+\c_{3,13}+\c_{3,17}+\c_{3,23}|\*2
   +|\c_{4,1 }+\c_{4,11}+\c_{4,19}+\c_{4,29}|\*2\cr
   &+[(\c_{1,7}+\c_{1,13}+\c_{1,17}+\c_{1,23})(\c_{3,1 }
   +\c_{3,11}+\c_{3,19}+\c_{3,29})\** +\;cc]\cr
   & +[(\c_{4,7}+\c_{4,13}+\c_{4,17}+\c_{4,23})(\c_{2,1 }
   +\c_{2,11}+\c_{2,19}+\c_{2,29})\** +\;cc]; &(A.8)\cr
\EE''_{8,28}\equi &|\c_{1,1}+\c_{1,11}+\c_{1,19}+\c_{1,29}+
                 \c_{9,1}+\c_{9,11}+\c_{9,19}+\c_{9,29}\cr
               &+\c_{5,7}+\c_{5,13}+\c_{5,17}+\c_{5,23}|\*2
                +| \c_{5,1}+\c_{5,11}+\c_{5,19}+\c_{5,29}\cr
               &+ \c_{3,7}+\c_{3,13}+\c_{3,17}+\c_{3,23}
           + \c_{7,7}+\c_{7,13}+\c_{7,17}+\c_{7,23}|\*2 \; ; & (A.9)\cr
}$$
where the characters $\c_{a,b}$ are just products of the characters of
$A_1\*{(1)}$: $\c_{a,b}=\c_a\c_b$.

The invariants (A.6) and (A.8) were already found in ref.~\FKSV. The other two
invariants have not appeared in the literature before: the invariant (A.7) can
be obtained from the $A_{1,3}A_{1,8}A_{1,1}\subset F_{4,1}$ conformal
embedding, with $A_{1,1}$ projected out, while the other, (A.9), arises from a
$A_{1,8}A_{1,28}\subset F_{4,1}$ conformal embedding.

Together with the results given in ref.~\HK{} for $k_1=k_2$, the above
equations exhaust all the $A_1+A_1$ physical invariants for $k_1,k_2\leq 21$
and $k_1=28$, $k_2\leq 21$. These include all of the levels where the arguments
of \SUSU{} break down, so this completes the $A_{1,k_1}+A_{1,k_2}$
classification for all levels: the complete list of all physical invariants is
given by the above invariants, along with the exceptionals $(4.3f)$, $(4.3g)$,
$(4.3h$), $(4.3i$) in \HK, the exceptionals $\EE_{10}\o\EE_{10}$,
$\EE_{16}\o\EE_{16}$, and $\EE_{28}\o \EE_{28}$, and the {\it conjugations}
\HK{} of all these when $k_1=k_2$.

\vfill\eject
\centerline {Table 1. The candidate types}
\medskip
$$\vbox{\tabskip=0pt\offinterlineskip
  \def\tablerule{\noalign{\hrule}}
  \halign to 5.75in{
    \strut#&\vrule#\tabskip=0em plus2em &    
    \hfil#&\vrule#&\hfil#&\vrule#&    
    \hfil#&\vrule#&\hfil# &\vrule#    
    \tabskip=0pt\cr\tablerule
&&\omit\hidewidth n\hidewidth&&
 \omit\hidewidth ${\cal T}\*+=({\cal T}_L\*+;{\cal T}_R\*+)$&&
\omit\hidewidth ${\cal T}\*0=({\cal T}_{L}\*0;{\cal T}_{R}\*0)$\hidewidth&&
\omit\hidewidth $\ell$\hidewidth&\cr\tablerule
&& 1 && $(B_{4,14}\,;\,-)$&& -- &&--& \cr
&& 2 && $(C_{4,10}\,;\,-)$&& -- &&--& \cr
&& 3 && $(D_{4,36}\,;\,-)$&& $(C_{2}\*2\,;\,-)$&&--& \cr
&& 4 && $(F_{4,108}\,;\,-)$&& $(C_{2}\*2\,;\,-)$&&29& \cr
&& 5  && $(G_{2,11}G_{2,206}\,;\,-)$&& $(C_{2}\*2\,;\,-)$&&13& \cr
&& 6 && $(G_{2,12}G_{2,108}\,;\,-)$&&$(-\,;\,G_{2}\*2)$&&13& \cr
&& 7  && $(G_{2,14}G_{2,59}\,;\,-)$&& $(C_{2}\*2\,;\,-)$&&11& \cr
&& 8  && $(G_{2,17}G_{2,38}\,;\,-)$&& $(C_{2}\*2\,;\,-)$&&13& \cr
&& 9 && $(G_{2,24}G_{2,24}\,;\,-)$&&$(-\,;\,G_{2}\*2)$&&--& \cr
&& 10 && $(A_{2,1}\,;\,A_{1,1}A_{1,1})$&& $(D_{7}A_{1}\,;\,-)$&&5& \cr
&& 11 && $(A_{2,3}\,;\,A_{1,2}A_{1,10})$&&$(C_2A_1\,;\,A_3)$&&--& \cr
&& 12 && $(A_{2,3}\,;\,A_{1,4}A_{1,4})$&&$(-\,;\,A_2)$&&--& \cr
&& 13 && $(A_{2,4}\,;\,A_{1,4}A_{1,12})$&&$(C_2\,;\,A_6)$&&--& \cr
&& 14 && $(A_{2,5}\,;\,A_{1,5}A_{1,40})$&&$(C_3\,;\,C_2\*2A_3)$&&11& \cr
&& 15 && $(A_{2,5}\,;\,A_{1,6}A_{1,22})$&&$(-\,;\,A_2G_2\*3)$&&5& \cr
&& 16 && $(A_{2,5}\,;\,A_{1,7}A_{1,16})$&&$(D_5D_7\,;\,A_3A_1)$&&5& \cr
&& 17 && $(A_{2,5}\,;\,A_{1,8}A_{1,13})$&&$(D_5\,;\,C_2\*2A_1)$&&11& \cr
&& 18 && $(A_{2,5}\,;\,A_{1,10}A_{1,10})$&&$(-\,;\,A_2)$&&7& \cr
&& 19 && $(A_{2,6}\,;\,A_{1,8}A_{1,88})$&&$(-\,;\,A_2)$&&7& \cr
&& 20 && $(A_{2,6}\,;\,A_{1,10}A_{1,34})$&&$(A_1\,;\,G_2\*2C_2A_3)$&&5& \cr
&& 21 && $(A_{2,6}\,;\,A_{1,16}A_{1,16})$&&$(-\,;\,A_2)$&&7& \cr
&& 22 && $(A_{2,7}\,;\,A_{1,14}A_{1,238})$&&$(-\,;\,A_{14}C_2G_2)$&&7& \cr
&& 23 && $(A_{2,7}\,;\,A_{1,16}A_{1,88})$&&$(C_2\*2A_{14}\,;\,-)$&&11& \cr
&& 24 && $(A_{2,7}\,;\,A_{1,18}A_{1,58})$&&$(C_2A_1\,;\,A_3)$&&11& \cr
&& 25 && $(A_{2,7}\,;\,A_{1,22}A_{1,38})$&&$(-\,;\,A_{14}C_2G_2)$&&7& \cr
&& 26 && $(A_{2,7}\,;\,A_{1,28}A_{1,28})$&&$(A_1\,;\,B_3A_{14})$&&7& \cr
&& 27 && $(A_{2,8}\,;\,A_{1,32}A_{1,1120})$&&$(-\,;\,A_{10}C_2\*2)$&&13& \cr
&& 28 && $(A_{2,8}\,;\,A_{1,34}A_{1,394})$&&$(-\,;\,A_{10}C_2\*2G_2\*3)$&&5&
\cr
&& 29 && $(A_{2,8}\,;\,A_{1,40}A_{1,152})$&&$(-\,;\,A_{10}C_2\*2)$&&13& \cr
&& 30 && $(A_{2,8}\,;\,A_{1,42}A_{1,130})$&&$(A_3A_1\,;\,C_2\*2)$&&17& \cr
&& 31 && $(A_{2,8}\,;\,A_{1,64}A_{1,64})$&&$(-\,;\,A_2)$&&5& \cr
\tablerule\noalign{\smallskip}
 }} $$
\centerline{$n=$ label; ${\cal T}\*+=$ positive type;
${\cal T}\*0=$ null augment (level 0 subscripts}
\centerline{ omitted); and
$\ell=$ first number (if any) which violates $\rho$ parity test.}\vfill\eject

\centerline {Table 1 (cont.). The candidate types}
\medskip
$$\vbox{\tabskip=0pt\offinterlineskip
  \def\tablerule{\noalign{\hrule}}
  \halign to 5.75in{
    \strut#&\vrule#\tabskip=0em plus2em &    
    \hfil#&\vrule#&\hfil#&\vrule#&    
    \hfil#&\vrule#&\hfil# &\vrule#   
    \tabskip=0pt\cr\tablerule
&&\omit\hidewidth n\hidewidth&&
 \omit\hidewidth ${\cal T}\*+=({\cal T}_L\*+;{\cal T}_R\*+)$
\hidewidth&&  \omit\hidewidth ${\cal T}\*0=({\cal T}_{L}\*0;
{\cal T}_{R}\*0)$\hidewidth&&  \omit\hidewidth $\ell$\hidewidth&\cr\tablerule
&&32  && $(C_{2,1}\,;\,A_{1,1}A_{1,2})$&&$(C_2\,;\,G_2)$&&5& \cr
&&33  && $(C_{2,2}\,;\,A_{1,2}A_{1,10})$&&$(C_2\,;\,G_2)$&&7& \cr
&&34  && $(C_{2,2}\,;\,A_{1,4}A_{1,4})$&&$(-\,;\,C_2)$&&7& \cr
&&35  && $(C_{2,3}\,;\,A_{1,5}A_{1,40})$&&$(G_2A_1\,;\,C_3)$&&17& \cr
&&36  && $(C_{2,3}\,;\,A_{1,6}A_{1,22})$&&$(C_2\,;\,G_2)$&&5& \cr
&&37  && $(C_{2,3}\,;\,A_{1,7}A_{1,16})$&&$(-\,;\,C_3G_2\*2A_1)$&&5& \cr
&&38 && $(C_{2,3}\,;\,A_{1,8}A_{1,13})$&&$(G_2A_1\,;\,C_3)$&&7& \cr
&&39  && $(C_{2,3}\,;\,A_{1,10}A_{1,10})$&&$(G_2\*2C_2\,;\,-)$&&--& \cr
&&40  && $(C_{2,4}\,;\,A_{1,20}A_{1,460})$&&$(A_2\,;\,A_6)$&&13& \cr
&&41  && $(C_{2,4}\,;\,A_{1,22}A_{1,166})$&&$(A_3A_1\,;\,C_2A_6)$&&5& \cr
&&42  && $(C_{2,4}\,;\,A_{1,26}A_{1,82})$&&$(G_2\,;\,C_6)$&&5& \cr
&&43  && $(C_{2,4}\,;\,A_{1,28}A_{1,68})$&&$(A_2\,;\,A_6)$&&11& \cr
&&44  && $(C_{2,4}\,;\,A_{1,40}A_{1,40})$&&$(-\,;\,C_2)$&&--& \cr
&&45  && $(G_{2,1}\,;\,A_{1,1}A_{1,3})$&&$(C_2A_1\,;\,B_3)$&&--& \cr
&&46  && $(G_{2,2}\,;\,A_{1,3}A_{1,43})$&&$(C_2\*2G_2\,;\,-)$&&13& \cr
&&47  && $(G_{2,2}\,;\,A_{1,4}A_{1,16})$&&$(-\,;\,C_2\*2)$&&5& \cr
&&48  && $(G_{2,2}\,;\,A_{1,7}A_{1,7})$&&$(D_7\,;\,A_3)$&&7& \cr
&&49  && $(A_{2,3}\,;\,C_{2,2})$&&$(A_2\,;\,-)$&&7& \cr
&&50  && $(A_{2,5}\,;\,C_{2,3})$&&$(-\,;\,D_7A_1)$&&7& \cr
&&51  && $(A_{2,15}\,;\,C_{2,6})$&&$(A_2\,;\,-)$&&--& \cr
&&52  && $(A_{2,21}\,;\,C_{2,7})$&&$(C_2\,;\,A_2)$&&11& \cr
&&53  && $(A_{2,30}\,;\,C_{2,8})$&&$(A_2\,;\,-)$&&5& \cr
&&54  && $(A_{2,45}\,;\,C_{2,9})$&&$(A_2\,;\,-)$&&5& \cr
&&55  && $(A_{2,75}\,;\,C_{2,10})$&&$(A_2\,;\,-)$&&5& \cr
&&56  && $(A_{2,165}\,;\,C_{2,11})$&&$(A_2A_3\*2\,;\,C_6)$&&5& \cr
&&57  && $(A_{2,9}\,;\,G_{2,3})$&&$(A_3A_1\,;\,-)$&&--& \cr
&&58  && $(A_{2,21}\,;\,G_{2,4})$&&$(A_3A_1\,;\,-)$&&5& \cr
&&59  && $(A_{2,105}\,;\,G_{2,5})$&&$(A_3A_1\,;\,-)$&&--& \cr
&&60  && $(C_{2,7}\,;\,G_{2,4})$&&$(A_3\,;\,A_2A_1)$&&--& \cr
&&61  && $(C_{2,42}\,;\,G_{2,8})$&&$(G_2\*3C_2\,;\,-)$&&7& \cr
&&62  && $(A_{1,4}A_{1,4};A_{1,2}A_{1,10})$&&$(G_2\*3\,;\,-)$&&--& \cr
\tablerule\noalign{\smallskip}
 }} $$
\centerline{$n=$ label; ${\cal T}\*+=$ positive type;
${\cal T}\*0=$ null augment (level 0 subscripts}
\centerline{ omitted); and
$\ell=$ first number (if any) which violates $\rho$ parity test.}\vfill\eject

\centerline {Table 2. The Parity Test Survivors}
\medskip
$$\vbox{\tabskip=0pt\offinterlineskip
  \def\tablerule{\noalign{\hrule}}
  \halign to 5.75in{
    \strut#&\vrule#\tabskip=0em plus2em &    
    \hfil#&\vrule#&\hfil#&\vrule#&    
    \hfil#&\vrule#&\hfil# &\vrule#   
    \tabskip=0pt\cr\tablerule
&&\omit\hidewidth n\hidewidth&&
 \omit\hidewidth P
\hidewidth&&  \omit\hidewidth Dim
\hidewidth&&  \omit\hidewidth Method \hidewidth&\cr\tablerule
&& 1 && 0 && 3 && Lattice (3) & \cr
&& 2 && 1 && 2 && Lattice (3) & \cr
&& 3 && 0 && 9 && Lattice (3) & \cr
&& 9 && 0 && - && Projection (5.2) & \cr
&& 11 && 1 && 1 && Explicit (5.1) & \cr
&& 12 && 0 && 0 && Explicit (5.1) & \cr
&& 13 && 1 && 1 && Explicit (5.1) & \cr
&& 39 && 1 && - && Cardinality (4.2) & \cr
&& 44 && 0 && 4 && Lattice (3) & \cr
&& 45 && 1 && 1 && Explicit (5.1) & \cr
&& 51 && 0 && 0 && Lattice (3) & \cr
&& 57 && 1 && - && Cardinality (4.2) & \cr
&& 59 && 0 && - && Cardinality (4.1) & \cr
&& 60 && 1 && - && Cardinality (4.2) & \cr
&& 62 && 0 && 0 && Explicit (5.1) & \cr
\tablerule\noalign{\smallskip}
 }} $$
\centerline{$n=$ label; P=number of physical invariants;
Dim=dimension of commutant}
\centerline{(if known); and
Method is technique used (relevant Section in brackets).}\vfill\eject

\bigskip\bigskip
\noindent{{\bf References}} \bigskip
\frenchspacing
\item{\CIZ.} A. Cappelli, C. Itzykson and  J.-B. Zuber,
Nucl. Phys. B280 [FS18] (1987) 445
\item{\MS.} G. Moore and N. Seiberg, Nucl. Phys. B313 (1989) 16
\i{\SY.} A.~N. Schellekens and S. Yankielowicz, Nucl. Phys. B327
(1989) 673;
\i{} A.~N. Schellekens, Phys. Lett. B244 (1990) 255;
\i{} B. Gato-Rivera and A.~N. Schellekens, Commun. Math. Phys. 145 (1992) 85
\item{\ITZ.} C. Itzykson, Nucl. Phys. (Proc. Suppl.) 5B
(1988) 150;
\item{} P. Degiovanni, Commun. Math. Phys. 127 (1990) 71
\item{\COMM.} T. Gannon, Nucl. Phys. B396 (1993) 708
\i{\KA.} A. Kato, Mod. Phys. Lett. A2 (1987) 585;
\item{} A. Cappelli, C. Itzykson and  J.-B. Zuber,
Commun. Math. Phys. 113 (1987) 1;
\item{} D. Gepner and Z. Qiu, Nucl. Phys. B285 (1987) 423
\i{\Kas.} D. Kastor, Nucl. Phys. B280 [FS18] (1987) 304
\i{\GA.} T. Gannon, ``The classification of affine SU(3) modular invariant
partition functions'', Commun. Math. Phys. (to appear)
\i{\SUSU.} T. Gannon, ``Towards a classification of su(2)$\oplus\cdots\oplus$
su(2) modular invariant partition functions'' (in preparation)
\i{\CL.} G.~B. Cleaver and D.~C. Lewellen, ``On modular invariant partition
functions for tensor products of conformal field theories'', (Santa Barbara
ITP preprint NSF-ITP-92-148, 1992)
\item{\HET.} T. Gannon, Nucl. Phys. B402 (1993) 729
\item{\WA.} N.~P. Warner, Commun. Math. Phys. 130 (1990) 205;
\item{} P. Roberts, Phys. Lett. B244 (1990) 429;
\item{} P. Roberts and H. Terao, Int. J. Mod. Phys. A7 (1992) 2207
\i{\HK.} T. Gannon and Q. Ho-Kim, ``Low level modular invariants of rank two
algebras'', Int. J. Mod. Phys. A (to appear)
\item{\CS.} J.~H. Conway and N.~J.~A. Sloane, {\it Sphere Packings,
Lattices and Groups}, (Springer, New York, 1988).
\i{\Gal.} T. Gannon and C.~S. Lam, Rev. Math. Phys. 3 (1991) 331
\item{\RTW.} Ph. Ruelle, E. Thiran and J. Weyers, Nucl. Phys. B402 (1993) 693
\item{\KAC.} V.~G. Kac, {\it Infinite Dimensional Lie Algebras}, 3rd ed.,
(Cambridge University Press, Cambridge, 1990);
\item{} S. Kass, R.~V. Moody, J. Patera and R. Slansky, {\it Affine Lie
Algebras, Weight Multiplicities, and Branching Rules} Vol.1 (University
of California Press, Berkeley, 1990)
\i{\CG.} A. Coste and T. Gannon, ``Remarks on Galois symmetry in RCFT'',
{\it Phys. Lett.} {\bf B} (to appear).
\i{\FKSV.} J. Fuchs, A. Klemm, M. Schmidt and D. Verstegen, Int. J. Mod.
Phys. A7 (1992) 2245;
\i{} D. Verstegen, Nucl. Phys. B346 (1990) 349
\i{\CE.} S. Bais and P. Bouwknegt, Nucl. Phys. B279 (1987) 561;
\i{} A.~N. Schellekens and N.~P. Warner, Phys. Rev. D34 (1986) 3092
\i{\BOW.} P. Bouwknegt, Nucl. Phys. B290 [FS20] (1987) 507
\i{\SCH.} A.~N. Schellekens, Commun. Math. Phys. 153 (1993) 159

\end